\documentclass[prl,a4paper,twocolumn,showpacs,floatfix,superscriptaddress,amsmath,amsfonts,amssymb,preprintnumbers,citeautoscript]{revtex4}
\pdfoutput=1
\usepackage[T1]{fontenc}\usepackage[latin1]{inputenc}
\usepackage[pdftex, dvipsnames]{xcolor}
\usepackage{dcolumn,graphicx,color,booktabs,microtype,afterpage}
\graphicspath{{Figures/}}
\usepackage[charter,greekuppercase=italicized]{mathdesign}
\usepackage{sidecap}\usepackage{physics}
\renewcommand{\figurename}{Fig.}
\renewcommand{\tablename}{Table}
\makeatletter\renewcommand{\fnum@figure}[1]{\figurename~\thefigure.}\makeatother
\makeatletter\renewcommand{\fnum@table}[1]{\tablename~\thetable.}\makeatother

\newcount\hh \newcount\mm
\hh=\time \divide\hh by 60
\mm=\hh \multiply\mm by 60 \mm=-\mm
\advance\mm by \time
\def\now{\number\hh:\ifnum\mm<10{}0\fi\number\mm}

\usepackage[colorlinks,plainpages=false,linkcolor=blue,urlcolor=blue,citecolor=blue,pdfpagemode=UseNone,pdfstartview=FitBH]{hyperref}

\begin{document}

\makeatletter\renewcommand{\ps@plain}{
\def\@evenhead{\hfill\itshape\rightmark}
\def\@oddhead{\itshape\leftmark\hfill}
\renewcommand{\@evenfoot}{\hfill\small{--~\thepage~--}\hfill}
\renewcommand{\@oddfoot}{\hfill\small{--~\thepage~--}\hfill}
}\makeatother\pagestyle{plain}


\title{Weak doping dependence of the antiferromagnetic coupling\\between nearest-neighbor Mn$^{2+}$ spins in (Ba$_{1-x}$K$_{x}$)(Zn$_{1-y}$Mn$_{y}$)$_{2}$As$_{2}$}

\author{M.~A.~Surmach}
\affiliation{Institut f{\"u}r Festk{\"o}rper- und Materialphysik, TU Dresden, D-01069 Dresden, Germany}
\author {B.~J.~Chen}
\affiliation{Institute of Physics, Chinese Academy of Sciences, 100190 Beijing, China}
\affiliation{Center for High Pressure Science and Technology Advanced Research, Beijing 100094, China}
\author{Z.~Deng}
\affiliation{Institute of Physics, Chinese Academy of Sciences, 100190 Beijing, China}
\author{C.~Q.~Jin}
\affiliation{Institute of Physics, Chinese Academy of Sciences, 100190 Beijing, China}
\affiliation{School of Physics, University of Chinese Academy of Sciences, Beijing 100049, China}
\affiliation{Collaborative Innovation Center of Quantum Matter, Beijing, China}
\author{J.~K.~Glasbrenner}
\affiliation{Department of Computational and Data Sciences/Center for Simulation and Modeling, George Mason University, 4400 University Drive, Fairfax, VA 22030}
\affiliation{Code 6393, Naval Research Laboratory, Washington, DC 20375, USA}
\author{I.~I.~Mazin}
\affiliation{Code 6393, Naval Research Laboratory, Washington, DC 20375, USA}
\author{A.~Ivanov}
\affiliation{Institut Laue-Langevin, 71 avenue des Martyrs, CS 20156, F-38042 Grenoble Cedex 9, France}
\author{D.~S.~Inosov}\email[Corresponding author: \vspace{5pt}]{Dmytro.Inosov@tu-dresden.de}
\affiliation{Institut f{\"u}r Festk{\"o}rper- und Materialphysik, TU Dresden, D-01069 Dresden, Germany}

\begin{abstract}
\parfillskip=0pt\relax
\noindent Dilute magnetic semiconductors (DMS) are nonmagnetic semiconductors doped with magnetic transition metals. The recently discovered DMS material (Ba$_{1-x}$K$_{x}$)(Zn$_{1-y}$Mn$_{y}$)$_{2}$As$_{2}$ offers a unique and versatile control of the Curie temperature, $T_{\mathrm{C}}$, by decoupling the spin (Mn$^{2+}$, $S=5/2$) and charge (K$^{+}$) doping in different crystallographic layers. In an attempt to describe from first-principles calculations the role of hole doping in stabilizing ferromagnetic order, it was recently suggested that the antiferromagnetic exchange coupling $J$ between the nearest-neighbor Mn ions would experience a nearly twofold suppression upon doping 20\% of holes by potassium substitution. At the same time, further-neighbor interactions become increasingly ferromagnetic upon doping, leading to a rapid increase of $T_{\mathrm{C}}$. Using inelastic neutron scattering, we have observed a localized magnetic excitation at about 13\,meV, associated with the destruction of the nearest-neighbor Mn\,--\,Mn singlet ground state. Hole doping results in a notable broadening of this peak, evidencing significant particle-hole damping, but with only a minor change in the peak position. We argue that this unexpected result can be explained by a combined effect of superexchange and double-exchange interactions.
\end{abstract}

\maketitle\enlargethispage{4.2pt}

\section{I. Introduction}\vspace{-1pt}

Magnetic semiconductors have attracted much attention in recent years as they combine ferromagnetism, required for spintronic applications, with semiconducting properties, deriving from materials used in conventional microelectronics \cite{XiaGeChang12}. Such a useful combination can be achieved by the substitution of host cations with magnetic ions. An intense research effort in 1970s\,--\,1980s has shown that a low concentration of magnetic impurities can introduce large magnetic effects with no degeneration of optical or transport properties \cite{Furdyna88}.

During the last two decades, a lot of effort has been invested to reveal the microscopic mechanism of ferromagnetism in dilute magnetic semiconductors (DMS), which are also termed functional ferromagnets due to the unique tunability of their magnetic properties \cite{Dietl03}. These materials derive from traditional semiconductors as they are doped with a small amount of localized magnetic impurities in addition to (or instead of) the conventional hole (p-type) or electron (n-type) doping. Scientific interest to DMS is driven by the need to increase their Curie temperature, $T_{\mathrm{C}}$, beyond room temperature to allow for applications in spintronic devices. After these efforts prove successful, these devices could be easily integrated with the conventional semiconductor technology.

The development of DMS-based spintronic devices is strongly connected with the efficiency of injection, transfer, and detection of spin-polarized currents from a ferromagnetic (FM) material into a semiconductor \cite{AktasMikailzade13}. However, the key obstacle for realizing this using a combination of a conventional metallic ferromagnet (such as iron or cobalt) and a semiconductor is the resistance mismatch at the metal/semiconductor interface, hindering an effective spin injection \cite{Jain1991}. DMS were recognized as a possible way around this problem after Dietl and colleagues published a theoretical work predicting room-temperature ferromagnetism in Mn-doped ZnO \cite{DietlOhno00}. Though some early experimental works reported FM ordering of 3$d$ TM ions in ZnO at room temperature \cite{SatoKatayama00}, these initial results could not be confirmed by other groups \cite{Dietl10}. In all prototypical DMS materials of the III-V and II-VI groups, such as Mn-doped GaAs or ZnO, maximal $T_{\mathrm{C}}$ remains limited to only 180\,--\,185\,K \cite{OlejnikOwen08, WangCampion08}. In the II-VI semiconductors, the isovalent substitution of Mn leads to the lack of carriers for the rise of a robust ferromagnetism, whereas the dual role of Mn in terms of both spin and charge doping in the III-V family complicates our theoretical understanding and restricts possible ways to enhance $T_{\mathrm{C}}$.

Only recently, these difficulties could be overcome in the I-II-V \cite{DengJin11, DengZhao13} and II-II-V \cite{ZhaoDeng13, ZhaoChen14, TaoLin15, SunLi16, SunZhao17} families of semiconductors doped with Mn$^{2+}$ ions, resulting in the new DMS materials Li$_{1+x}$(Zn$_{1-y}$Mn$_{y}$)As and (Ba$_{1-x}$K$_{x}$)(Zn$_{1-y}$Mn$_{y}$)$_{2}$As$_{2}$. In these two systems, hole or electron doping is decoupled from the spin injection, as they occur in different crystallographic layers, thus offering a unique possibility to tune the carrier concentration and the amount of magnetic moments independently. The maximal $T_{\text{C}}$ reported for (Ba$_{1-x}$K$_{x}$)(Zn$_{1-y}$Mn$_{y}$)$_{2}$As$_{2}$ reached 230\,K \cite{ZhaoChen14}, reviving the interest to the DMS problem. Moreover, this system is isostructural to the layered BaFe$_{2}$As$_{2}$ parent compound of the best-studied ``122'' family of iron-based superconductors \cite{Dai15, Inosov16}. This structure offers a versatility of chemically tailored properties ranging from metallic to semiconducting behavior and from anti- to ferromagnetism, which makes (Ba$_{1-x}$K$_{x}$)(Zn$_{1-y}$Mn$_{y}$)$_{2}$As$_{2}$ a promising model compound for building prototypes of future integrated spintronic devices.

The central open question, however, is the theoretical understanding of the doping-enhanced ferromagnetism in these new systems. A reliable theory is needed to guide the search for materials with even higher Curie temperatures, yet the ability to predict $T_{\mathrm{C}}$ quantitatively from first-principle calculations still remains elusive. According to a recent theoretical work \cite{GlasbrennerZutic14}, the ferromagnetism of localized Mn spins mediated by the itinerant As holes in (Ba$_{1-x}$K$_{x}$)(Zn$_{1-y}$Mn$_{y}$)$_{2}$As$_{2}$ arises from a competition between the short-range antiferromagnetic (AFM) superexchange interaction between the nearest-neighbor Mn$^{2+}$ ions and a longer-range FM effective double-exchange interaction for all other Mn-Mn distances. At low Mn concentrations, whenever two Mn ions occupy nearest-neighbor sites on the lattice, they form a singlet state due to the strong AFM interaction within the dimer, which effectively inactivates the AFM exchange channel. On the other hand, the remaining solitary Mn spins interact ferromagnetically, and the strength of this interaction enhances with hole doping, leading to a rapid stabilization of the FM order.

This theoretical picture was substantiated by the recent x-ray magnetic circular dichroism (XMCD) measurements at the As $K$~edge under ambient- and high-pressure conditions \cite{SunLi16}, which confirmed that the long-range magnetic order in (Ba$_{1-x}$K$_{x}$)(Zn$_{1-y}$Mn$_{y}$)$_{2}$As$_{2}$ is mediated by the $p$ states of As through As\,4$p$\,--\,Mn\,3$d$ hybridization. It was also demonstrated that the magnetic ordering of the bulk sample is intimately connected with the polarization of hole carriers and their mobility. A more recent investigation \cite{SunZhao17}, which includes x-ray emission (XES) and absorption (XAS) spectroscopy at the Mn\,$K$ edge, in addition to XMCD measurements at the As~$K$ edge, was focused on the evolution of Mn\,3$d$ and As\,4$p$ states and their hybridization with the doped holes under pressure. Authors explained the $T_{\text{C}}$ enhancement for (Ba$_{1-x}$K$_{x}$)(Zn$_{1-y}$Mn$_{y}$)$_{2}$As$_{2}$ in the presence of hole doping by the increase in $p$-$d$ hybridization strength at the cost of a reduction in Mn local spin density, resulting in enhanced indirect exchange interactions between Mn ions and inducing magnetic polarization in the As~$4p$ states. This result confirmed the proposed theoretical picture that magnetic interactions and $T_{\text{C}}$ are tunable and depend on the position of Mn~3$d$ bands as well as the Hubbard $U$ splitting between spin-up and spin-down states.

The hybridization of localized Mn~3$d$ and itinerant As~4$p$ states gives rise to the magnetic coupling between the Mn$^{2+}$ ions, consisting of competing AFM superexchange and FM double exchange terms \cite{KossutGaj10, GlasbrennerZutic14}. The density functional theory (DFT) calculations of the exchange parameters \cite{GlasbrennerZutic14} suggest that in the absence of hole doping the nearest- and next-nearest-neighbor interactions between the Mn spins are AFM, whereas all further-neighbor interactions are negligibly weak. Upon hole doping, the calculated interactions become increasingly FM, so that for $x=20$\% doping a sizable long-range FM coupling appears. As regards the most compact Mn pair, where two Mn subsitute two nearest neighbors, the calculated ground state remains an AFM singlet, but the calculated energy cost of flipping one Mn spin to make a fully polarized FM dimer, $S=5$ FM state, is reduced by nearly a factor of two compared to the undoped case.

In the simplified picture of two spins coupled by both Heisenberg superexchange and canonical double exchange, one expects the energy of the lowest spin excitation (from $S=0$ to $S=\pm1)$ to be $\Delta{}E\left(\Delta{}S=1\right)=2J-\frac{1}{6}t_{\textrm{eff}}$ as explained in more details below. Motivated by this, we have measured the energy of this excitation using inelastic neutron scattering (INS), expecting to find the lowest excitation energy to be reduced by a factor of 2. Note that INS is a bulk-sensitive method with momentum resolution, in contrast to optical techniques or magnetic-resonance {studies~\cite{Furdyna88}}.

\vspace{-2pt}\section{II. Experimental results}\vspace{-2pt}

\begin{figure}[b]
\includegraphics[width=1.01\columnwidth]{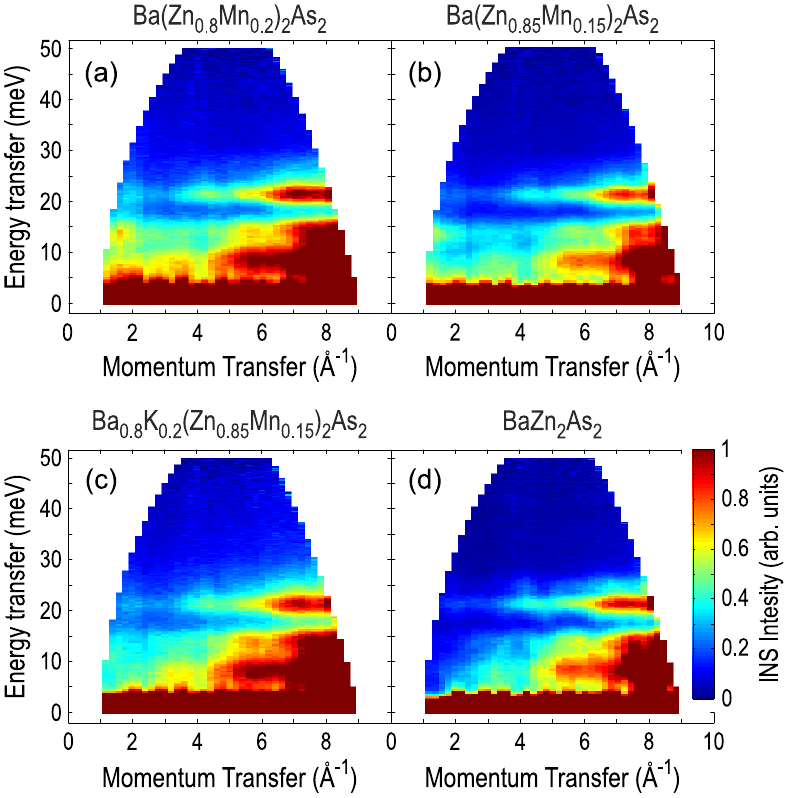}\vspace{2pt}
\caption{Intensity maps from TOF neutron spectroscopy (raw data) for II\,--\,II\,--\,V type DMS with isovalent Mn substitution: (a,b)~undoped Ba(Zn$_{1-y}$Mn$_{y}$)$_{2}$ with {$y=0.2$} and 0.15; (c)~hole-doped Ba$_{1-x}$K$_{x}$(Zn$_{1-y}$Mn$_{y}$)$_{2}$ with {$x=0.2$}, {$y=0.15$}; and (d) the nonmagnetic reference sample BaZn$_{2}$As$_{2}$. The intensity in each panel was corrected for detector efficiencies using a vanadium standard and normalized to the weight of each sample.\vspace{-3pt}}
\label{fig:maps}
\end{figure}

The energy of the lowest-lying INS magnetic excitation provides information about the nearest-neighbor exchange constant. This excitation is local and therefore dispersionless, hence it was sufficient to perform our INS measurements on polycrystalline samples. Powders with 4 different compositions were chosen: two undoped Ba(Zn$_{1-y}$Mn$_{y}$)$_{2}$As$_{2}$ samples with $y=0.2$ (9.9\,g) and $y=0.15$ (9.7\,g), one doped sample with $x=0.2$ and $y=0.15$ (9.1\,g), and a nonmagnetic reference sample of BaZn$_{2}$As$_{2}$ (7.9\,g). The numbers given in brackets are sample masses, to which the measured INS intensity was normalized. The polycrystalline samples were synthesized by the solid-state reaction method in a high-purity argon atmosphere as described in the earlier studies \cite{ZhaoDeng13, ZhaoChen14}. The crystal structure, phase purity, and lattice constants of the resulting powders were examined by x-ray powder diffraction with a Philips X'pert diffractometer with Cu~K$_{\alpha}$ radiation and by magnetic susceptibility using a superconducting quantum interference device (SQUID-VSM, Quantum Design) at temperatures ranging from 2 to 300~K. The characterization revealed phase-pure compositions isostructural to the ``122'' iron pnictides (space group $I4/mmm$). The lattice parameters were consistent with those reported previously \cite{TaoLin15, ZhaoDeng13}.

We performed INS measurements at the thermal disk-chopper time-of-light (TOF) powder spectrometer IN4C at ILL, Grenoble (France). This instrument is equipped with a large detector bank consisting of 300 position-sensitive $^{3}$He tubes and has a resolving power of $\Delta E/E_{\mathrm{i}}\approx 4$\,--\,6\% \cite{CicognaniMutka00}. We wrapped all samples in an aluminum foil and fixed them on holders that were placed inside the standard orange-type cryostat. The incident neutron wavelength $\lambda _{\text{i}}$ was fixed at 1.2\,\AA, corresponding to $E_{\mathrm{i}}=56.8$\,meV, and the energy resolution (defined as the full-width at half-maximum of the elastic line) was set to 3.5\,meV. All measurements were performed at the base temperature of $T=1.6$\,K.

\begin{figure}[t!]
\vspace{2pt} {\hspace{-0.007\columnwidth}
\includegraphics[width=1.02\columnwidth]{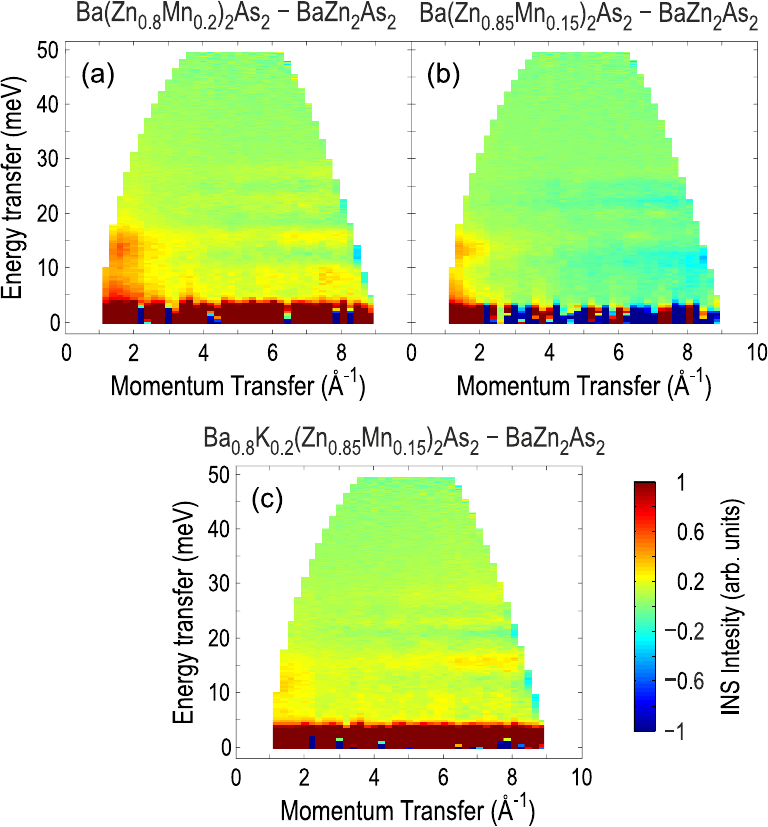}}
\caption{The same data as in Figs.~\ref{fig:maps}(a\,--\,c) after subtraction of the nonmagnetic background: (a,b)~undoped Ba(Zn$_{1-y}$Mn$_{y}$)$_{2}$ with {$y=0.2$} and 0.15; (c)~hole-doped Ba$_{1-x}$K$_{x}$(Zn$_{1-y}$Mn$_{y}$)$_{2}$ with {$x=0.2$}, {$y=0.15$}. In all panels, the signal from the nonmagnetic reference sample BaZn$_{2}$As$_{2}$ [shown in Fig.\,\ref{fig:maps}(c)] has been subtracted to reduce the phonon background and reveal the magnetic signal at low $|\mathbf{Q}|$, which is now clearly seen around 14\,meV. The horizontal features extending to higher $|\mathbf{Q}|$ are artifacts resulting from imperfect phonon subtraction.}
\label{fig:substr}
\end{figure}

We first present unprocessed INS data for each sample, shown as intensity maps in Fig.\,\ref{fig:maps} (panels a\,--\,d). The powder-averaged TOF data were corrected for possible inhomogeneities in detector efficiency using a vanadium standard and then normalized to the mass of each sample. The data were combined and transformed into energy-momentum space using the open-source software \textsc{Lamp} \cite{RichardFerrandKearley}. The data are dominated by several phonon lines that increase in intensity towards higher $\mathbf{Q}$. In addition, in panels (a)\,--\,(c) one can see a weaker magnetic signal at $|\mathbf{Q}|\approx1$\,--\,2\,\AA$^{-1}\!$, whose intensity decays with $\mathbf{Q}$ following the magnetic form factor. Relatively low magnetic intensity can be explained by the fact that Mn pairs form only a part of the total number of substituted Mn atoms.

\begin{figure}[t]
\includegraphics[width=\columnwidth]{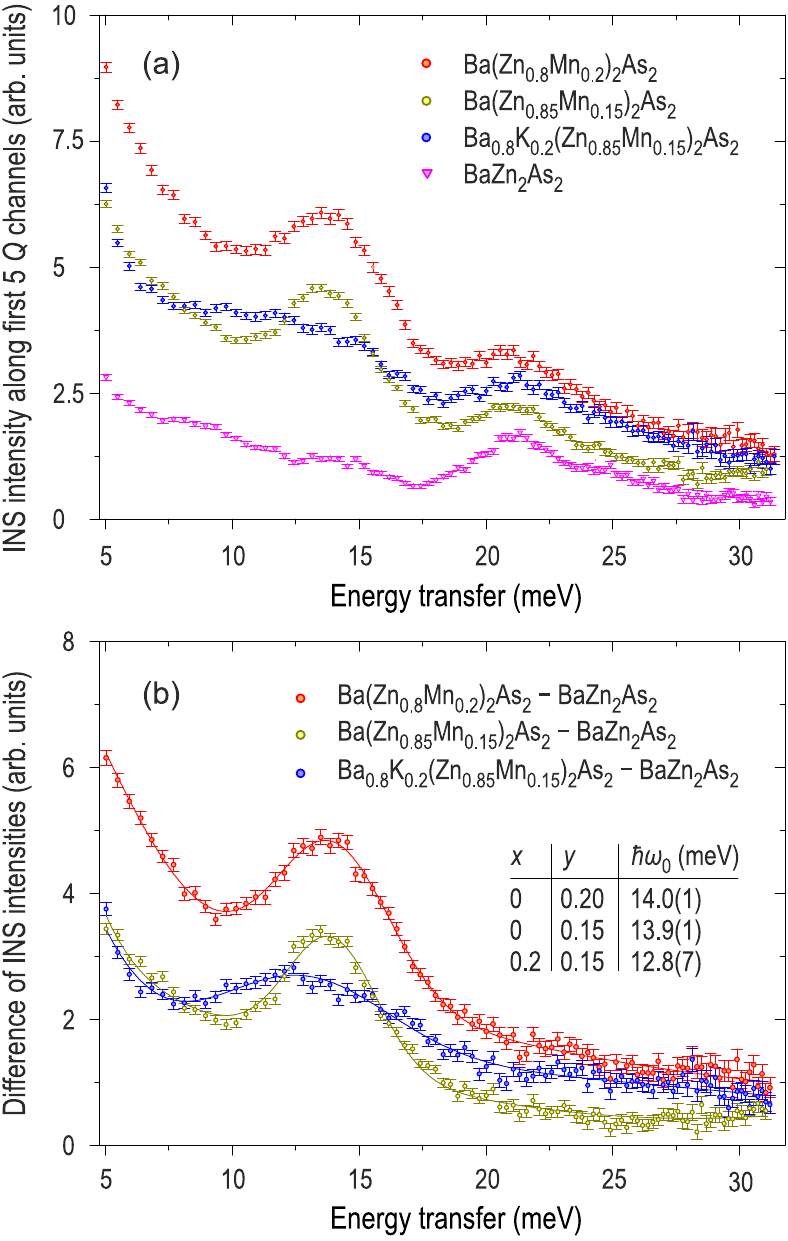}
\caption{(a)~Energy spectra obtained by averaging the first 5 momentum channels of the INS signals in Figs.~\ref{fig:maps}(a\,--\,d). (b)~The same for the subtracted signals in \ref{fig:substr}(a\,--\,c), which contain only the magnetic part of the signal. A clear peak originating from the lowest-energy $|\Delta S|=1$ excitation of the nearest-neighbor Mn dimers is seen around 13\,--\,14\,meV in all three samples. The fitted peak positions are summarized in the table inset.}
\label{fig:peaks}
\end{figure}

As the next step, in order to separate the magnetic signal from the nonmagnetic background and phonon contributions, we subtracted the nonmagnetic measurement of the reference compound BaZn$_{2}$As$_{2}$ from the measured intensity of each DMS sample. As all our compounds have the same lattice structure and similar lattice constants, their phonon spectra must also be similar, and the nonmagnetic contribution can be largely suppressed. The results of the subtraction are shown in Figs.~\ref{fig:substr}(a\,--\,c), where the same low-$|\mathbf{Q}|$ magnetic excitation is now clearly seen around 14\,meV. This energy scale is smaller by a factor of $\sim$\,1.9 than the theoretical prediction for the undoped compound \cite{GlasbrennerZutic14} (see Table~\ref{Tab:1}). Apart from the main magnetic peak corresponding to the nearest-neighbor Mn pairs, we see no other clear magnetic signals apart from the broad diffuse tail of intensity at low energies that does not exhibit any well-defined structure in energy and possibly originates from paramagnetic solitary Mn$^{2+}$ spins. This is consistent with the fact that for low concentrations of magnetic impurities, scattering from larger clusters (triads, etc.) is negligibly weak \cite{KepaKolesnik06}.

For a more quantitative analysis of the data, in Figs.~\ref{fig:peaks}(a) and \ref{fig:peaks}(b) we plotted the average of the first 5 momentum channels from Figs.~\ref{fig:maps} and \ref{fig:substr}, respectively. Figure~\ref{fig:peaks}(a) therefore shows the raw energy spectra for all samples, integrated in the low-$|\mathbf{Q}|$ region, whereas Fig.\,\ref{fig:peaks}(b) shows only magnetic parts of the same spectra after corresponding intensity subtractions. In all our three DMS samples, a single magnetic peak is seen on top of a broad incoherent background. The fitted peak positions are summarized in the table inset to Fig.\,\ref{fig:peaks}(b). First of all, one sees that for both undoped compounds the peak appears at the same energy of $\sim$14\,meV within the uncertainly of the measurements. This agrees well with the expectation that this energy is a local property of spin dimers and should be therefore independent of the Mn concentration \cite{GlasbrennerZutic14}. However, contrary to expectations, upon 20\% hole doping the peak position only slightly softens to 12.8\,meV, whereas the calculated energy for the full spin-flip is nearly twice smaller (Ref.\,\cite{GlasbrennerZutic14}, supplemental material).

At the same time, we observe a significant broadening of the peak upon potassium substitution as its full-width at half-maximum increases from 4.2\,meV in the undoped sample to 8.0\,meV in the 20\% hole-doped sample. After considering the instrumental resolution, this effect would correspond to a threefold increase in the intrinsic peak width. The most natural interpretation is that the As bands become metallic (and in fact spin-polarized) upon doping and thus generate Stoner continuum in the excitation spectrum. Coupling of the local Mn-dimer $S=0\to S=\pm1$ excitation to this continuum should naturally lead to the peak broadening. An additional source of broadening may come from the RKKY-mediated coupling of the $S=\pm1$ final state of such a transition to the solitary Mn$^{2+}$ spins and larger magnetic clusters, which similarly reduces the life time of the $\lvert\Delta{}S\rvert=1$ excitation as the carrier concentration increases.

\vspace{-2pt}\section{III. Discussion and summary}\vspace{-2pt}

\subsection{\vspace{-2pt}General considerations\vspace{-1pt}}

As we stated in the introduction, a simple picture for these exchange interactions is to adopt a two-spin model consisting of a Heisenberg superexchange term and a double-exchange term.
This two-site approximation can be justified by noting that, at low temperatures and low Mn concentrations, any nearest-neighbor Mn\,--\,Mn dimers will interact antiferromagnetically and form a singlet state of two spins $S_{1,2}=5/2$, while the remaining majority of Mn ions will not have close-by magnetic neighbors \cite{KepaKolesnik06}. As such, their superexchange interactions with other Mn$^{2+}$ ions can be neglected and will instead interact via double exchange. The standard approach has been to consider the superexchange and double exchange interactions separately and to assume that their combined effect is negligible, but in light of our results this may not be the case. Below we review the canonical results of the superexchange and double exchange models separately, after which we combine them and interpret the results.

The superexchange interaction for spin dimer is
\begin{equation}
H=2J\hat{\mathbf{S}}_{1}\cdot\hat{\mathbf{S}}_{2}.
\label{Eq:1}
\end{equation}
The magnetic state $\ket{S}$ of the dimer is characterized by its total spin $S$ defined as the maximal projection of the total spin operator $\hat{\mathbf{S}}=\hat{\mathbf{S}}_{1}\!+\hat{\mathbf{S}}_{2}$ for a given value of $\mathbf{S}^{2}=S(S+1)$. After rearranging Eq.\,(\ref{Eq:1}) using
\begin{equation}
2\hat{\mathbf{S}}_{1}\cdot\hat{\mathbf{S}}_{2}=\hat{\mathbf{S}}^{2}
-\hat{\mathbf{S}}^{2}_{1}-\hat{\mathbf{S}}^{2}_{2}
\end{equation}
and replacing all operators with their eigenvalues, we obtain the following set of eigenvalues of Eq.\,(\ref{Eq:1}):
\begin{equation}
E(S,S_{1},S_{2}) = J\left[  S(S\!+\!1)-S_{1}(S_{1}\!+\!1)-S_{2}(S_{2}\!+\!1)\right].
\label{Eq:3}
\end{equation}
For two identical particles, the energy difference between the final state $|S^{\prime}\rangle$ and the initial state $\ket{S}$ of the dimer is given~by
\begin{equation}
\Delta E(\Delta S) = J[S^{\prime}(S^{\prime}\!+1)-S(S+1)] = J\Delta S(2S+\Delta S+1).
\label{Eq:4}
\end{equation}
While the total spin of the dimer can assume all integer values between 0 and 5, neutron scattering is only able to probe excitations (transitions) with $\left\vert\Delta S\right\vert=1$ or $0$ in spin-flip and non-spin-flip processes, respectively \cite{Regnault06}. At low temperatures, when only the singlet state with $S=0$ is occupied, the only observable transition is that to the lowest $S^{\prime}=1$ excited state, with an energy $\Delta E=2J$. With increasing temperature, transitions from thermally populated initial states with higher spin become possible, resulting in equally spaced additional inelastic peaks on the energy-gain and energy-loss sides of the spectrum \cite{SvenssonHarvey78, FurrerGuedel79, GiebultowiczHolden88, GiebultowiczRhyne90}:
\begin{equation}
\Delta E(0,\pm1) = 0, \pm2J, \pm4J, \pm6J, \pm8J, \pm10J
\end{equation}

\begin{figure}[b]
\includegraphics[width=\columnwidth]{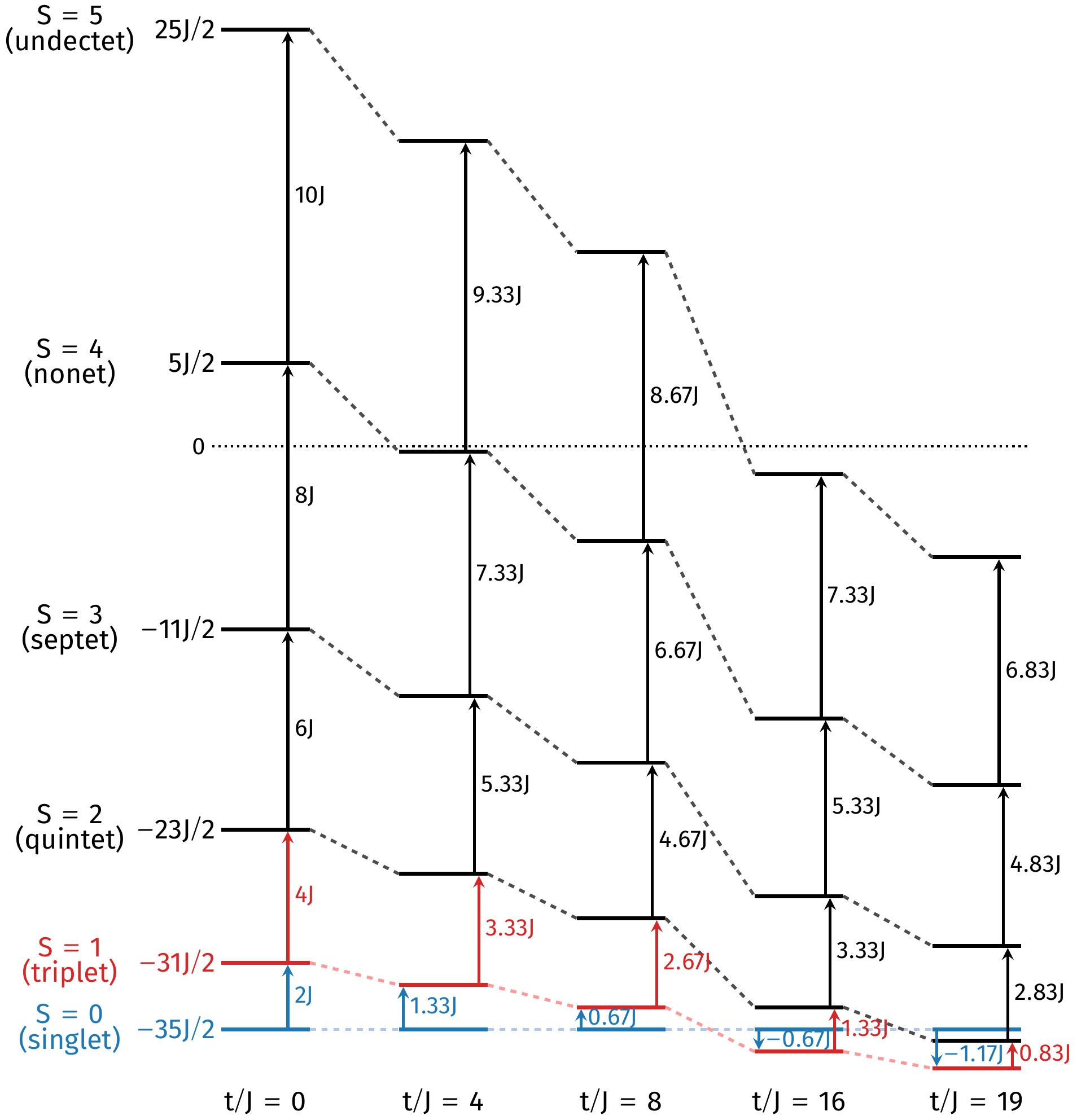}
\caption{Energy diagram of a single Mn\,--\,Mn dimer in dependence on the double exchange energy, parameterized here by the $t_{\textrm{eff}}/J$ ratio.\vspace{-1.5pt}}
\label{fig:se-de-peaks}
\end{figure}

The double exchange interaction is the result of an itinerant electron (or hole) interacting with localized spins (the Mn$^{2+}$ ions).
The Hamiltonian for a conduction electron interacting with two localized spin-$5/2$ sites is:
\begin{equation}
H_{\textrm{sf}} = -J_{\rm H} \sum_{i=1}^{2} \hat{\mathbf{s}}_{i} \cdot \hat{\mathbf{S}}_{i}
\label{Eq:sf-exchange}
\end{equation}
where $\hat{\mathbf{s}}$ and $\hat{\mathbf{S}}_{i}$ are the spin operators for the conduction election and localized sites, respectively. $J_{\rm H}$ is the local exchange (Hund's coupling) parameter, which for bad conductors is large compared to the conduction electron's effective hopping parameter $t_{\textrm{eff}}$, so $t_{\textrm{eff}} \ll J_{\rm H}$, leading to a strong preference for ferromagnetic alignment of the conduction electron and a nearby localized spin. One outcome is an energetic preference for the localized spins to develop a relative canting angle that depends on the total spin of the dimer~\cite{AndersonHasegawa55, NolthingRamakanth09},
\begin{equation}
\cos\vartheta{}/2=\frac{S_{0}+\frac{1}{2}}{2S_{1,2}+1},\textrm{ where }S_{0}=S\pm\frac{1}{2},
\end{equation}
which leads to the following ground state energy~\cite{AndersonHasegawa55, NolthingRamakanth09, Khomskii2010}:
\begin{equation}
E(S) = -t_{\textrm{eff}} \frac{S}{2S_{1,2} + 1},
\end{equation}
Here, $t_{\textrm{eff}}$ is the effective Mn\,--\,Mn hopping parameter, $S$ is the total spin of the dimer, and $S_{1,2}$ is the single particle spin. For spin-$5/2$ particles, this leads to double-exchange energies of $E = 0, -t/6, -t/3, -t/2, -2t/3, -5t/6$ for total spins of $S = 0, 1, 2, 3, 4, 5$, respectively. The analogous expression to Eq.\,(\ref{Eq:4}) for double exchange is:
\begin{equation}
\Delta{}E(\Delta{}S)=-t_{\textrm{eff}}\frac{\Delta{}S}{6}
\end{equation}
Combining this with Eq.\,(\ref{Eq:4}), we get our final expression:
\begin{equation}
\Delta{}E(\Delta{}S)=J\Delta{}S\left(2S+\Delta{}S+1\right)-t_{\textrm{eff}}\frac{\Delta{}S}{6}.
\label{eq:se-de-eigenvalues}
\end{equation}
Thus, the primary effect of double exchange is to reduce the $\lvert\Delta{}S\rvert=1$ transition energy by $t_{\textrm{eff}}/6$, independent of the total spin of the dimer.

\vspace{-2pt}\subsection{Model results}\vspace{-2pt}

The combined impact of superexchange and double exchange is controlled by the ratio $t_{\textrm{eff}}/J$. Let us treat this as an empirical parameter and consider how this shifts the superexchange energy levels of Eq.\,(\ref{Eq:3}) for several values of $t_{\textrm{eff}}/J$, see Fig.\,\ref{fig:se-de-peaks}. As $t_{\textrm{eff}}/J$ is increased from 0 to 19, the energy hierarchy changes twice, first at $t_{\textrm{eff}}/J = 12$ when the $S = 0$ (singlet) and $S = 1$ (triplet) states become degenerate and then at $t_{\text{eff}}/J = 18$ when $S = 0$ (singlet) and $S = 2$ (quintet) are degenerate. Despite these two changes to the energy hierarchy with increasing $t_{\text{eff}}/J$, the difference between $\Delta{}E(1\to 2)$ and $\Delta{}E(0\to 1)$ remains fixed at $2J$ for all values of $t_{\textrm{eff}}/J$.

Next, we estimate $J$ and $t_{\textrm{eff}}$ by fitting to the DFT energies reported in Ref.\,\cite{GlasbrennerZutic14}. To estimate $J$, we equate the DFT spin flip energy of $407$\,meV for zero hole doping with Eq.\,\ref{eq:se-de-eigenvalues}. We set $\Delta{}S = 5$, which is the analogue to a full spin flip, and $t_{\textrm{eff}}=0$, as there are no holes, and obtain $J = 13.6$\,meV. We follow a similar procedure to estimate $t_{\textrm{eff}}$ with the assumption that doping does not affect the superexchange coupling $J$. We equate the energy $207$\,meV for 20\% hole doping and set $\Delta{}S=1$ and obtain $t_{\textrm{eff}}=241$\,meV, corresponding to a ratio of $t_{\textrm{eff}}/J \approx 18$. This places (Ba, K)(Zn, Mn)$_{2}$As$_{2}$ with 20\% K doping on the far right side of the Fig.\,\ref{fig:se-de-peaks} diagram and at the $S = 0$ and $S = 2$ state degeneracy. We note that this degeneracy is accidental; the $t_{\textrm{eff}}/J = 18$ ratio also corresponds to the $S = 0 \to S = 1$ transition energy being reduced by a factor of two, thus matching the ratio between the DFT spin-flip energies for 0\% and 20\% K-doped systems. The predicted spectrum is reported in Table \ref{Tab:1}.

\begin{table}[b]
\begin{tabular}[c]{r@{~~~~~}r@{~~~~}r}
\toprule $\Delta S$ & $\Delta E (\textrm{SE})$ & $\Delta E (\textrm{SE}+\textrm{DE})$ \\
& (meV) & (meV)\\
\midrule 0 & 0 & 0 \\
1 & $-$13.0  & 27.2\\
2 & 14.2 & 54.4\\
3 & 41.4 & 81.6\\
4 & 68.6 & 108.8\\
5 & 95.8  & 136.0\\
\bottomrule
\end{tabular}
\caption{Calculated \mbox{$\Delta{}S=1$} transition energies using a super\-exchange-only model, $\Delta{}E(\textrm{SE})$, and a combined super\-exchange and double-exchange model, $\Delta{}E(\textrm{SE}+\textrm{DE})$, for $J = 13.6$\,meV and $t_{\textrm{eff}}=241$\,meV. The superexchange parameter $J$ and hopping parameter $t_{\textrm{eff}}$ were fitted to DFT spin-flip energies for 0\% and 20\% hole doping levels for Ba(Zn$_{1-y}$Mn$_{y}$)$_{2}$As$_{2}$ \cite{GlasbrennerZutic14}.\vspace{-4pt}}
\label{Tab:1}
\end{table}

\vspace{-2pt}\subsection{Comparison with the experimental data}\vspace{-2pt}

In the undoped samples, we have a well-defined rather narrow peak that is clearly associated with superexchange-coupled dimers. The calculated spin-flip energy, as discussed above, is about 80\% larger than what would be consistent with the experimental observation. This is not surprising: the superexchange interaction is proportional to $t_{\rm eff}^{2}/I$, where $t_{\rm eff}$ is the Mn\,--\,Mn effective hopping and $I$ is the energy cost of flipping the Mn spin. It is well known that if the Hubbard onsite repulsion is not fully accounted for, as was the case in the DFT calculations~of Ref.\,\cite{GlasbrennerZutic14}, this energy is underestimated by about $(U-J_{\rm H})/5$ \cite{PetukhovMazin03}, where $U$ is the Hubbard repulsion and $J_{\rm H}$ the Hund's rule coupling. This leads to an overestimation of the superexchange $J$, with a factor as large as 1.8 not being uncommon, similar to what we observe in Table \ref{Tab:1}.

More interestingly, the calculations indicate that the energy cost of a full spin flip is strongly reduced upon doping, while in the experiment the reduction is minor, but instead a huge broadening occurs. This makes us think that the Mn\,--\,Mn interactions in the doped sample are not described by the same two-spin Heisenberg Hamiltonian as in the undoped case, just with a reduced exchange constant. Using the double-exchange formalism leads to the ground state being weakly ferromagnetic, with the lowest $\Delta S=1$ transitions being $S=1\to S=0$ and $S=1\to S=2$ being practically degenerate and still too small when compared to the experiment.

Of course, while treating the superexchange interaction as a local one inside the dimer is fully justified, the double exchange is rather an interaction between a dimer and the itinerant, polarizable electron gas spanning over the entire crystal, that is to say, substantially nonlocal. The naive treatment presented in the previous section, in retrospect, appears inadequate. Rather, one should solve the entire problem in the spirit of the Fano model in optics, which leads to complex changes in the excitation line shape, but only to a very modest energy shift. The experimental data point out that such a theory would be more adequate, but it is, however, outside of the scope of our paper.

\section{IV. Acknowledgments}

We would like to thank Igor \v{Z}uti\'{c} from the University at Buffalo, NY, USA for stimulating discussions at the start of this project. We are also grateful to instrument scientists St\'{e}phane Rols and F\r{a}k Bj\"{o}rn from ILL for technical support during the experiment. This project was financially supported by the German Research Foundation (DFG) within the priority program SPP 1458/2 (Grant No.~IN209/1-2) and the Graduiertenkolleg GRK~1621 at the TU Dresden. Work at the Institute of Physics, Chinese Academy of Sciences, was supported by NSF and MOST of China through Research Projects. J.\,K.\,G.~acknowledges the support of the Office of Naval Research Summer Faculty Research Program. I.\,I.\,M. acknowledges funding from
the Office of Naval Research (ONR) through the Naval Research Laboratory's Basic Research Program.\vspace{-1ex}

\bibliographystyle{my-apsrev}
\bibliography{DMS}\vspace{-1ex}

\end{document}